\documentclass[aps,pra,reprint,amsmath,amssymb]{revtex4-1}

\usepackage{graphicx}
\usepackage{bm}
\usepackage{color}

\def\etal{{\it et~al.}}

\def\ohmcm{\,\Omega^{-1}{\rm cm^{-1}}}

\def\s1{\sigma_1(\omega)}

\begin{document}

\preprint{2015-11-13}

\title{\boldmath Terahertz ellipsometry study of the soft mode behavior in ultrathin SrTiO$_3$ films\unboldmath}

\author{P. Marsik}%
 \email{premysl.marsik@unifr.ch}
\author{K. Sen}
\author{J. Khmaladze}
\author{M. Yazdi-Rizi}
\author{B. P. P. Mallett}
\author{C. Bernhard}
\affiliation{University of Fribourg, Department of Physics and Fribourg Center for Nanomaterials,
Chemin du Mus\'{e}e 3, CH-1700 Fribourg, Switzerland}

\date{\today}

\begin{abstract}
We present a combined study with time-domain terahertz and conventional far-infrared ellipsometry of the temperature dependent optical response of SrTiO$_3$ thin films (82 and 8.5~nm) that are grown by pulsed-laser deposition on LSAT substrates. We demonstrate that terahertz ellipsometry is very sensitive to the optical response of these thin films, in particular, to the soft mode of SrTiO$_3$. We show that for the 82~nm film the eigenfrequency of the soft mode is strongly reduced by annealing at 1200~$^{\circ}$C, whereas for the 8.5~nm film it is hardly affected. For the latter, after annealing the mode remains at 125~cm$^{-1}$ at 300 K and exhibits only a weak softening to about 90~cm$^{-1}$ at 10~K. This suggests that this ultrathin film undergoes hardly any relaxation of the compressive strain due to the LSAT substrate. 
\end{abstract}



\maketitle

Perovskite titanates, particularly in the form of thin films, are important materials for device applications due to their high and tunable dielectric constant. Strontium Titanate (SrTiO$_3$, STO) is a quantum paraelectric material for which the ferroelectric transition is suppressed by the quantum fluctuations of the lattice.~\cite{Muller1979} The dielectric properties of STO in the static limit and at low frequencies are dominated by a strong polar phonon mode which is known as the soft mode since it exhibits a very strong softening at low temperature.~\cite{Petzelt2007} In bulk STO, the soft mode eigenfrequency is around 95~cm$^{-1}$ at room temperature and decreases to $\sim 10$~cm$^{-1}$ at low temperature. For epitaxial STO films, it was previously shown that this soft mode frequency, and therefore the dielectric constant, can differ substantially.~\cite{Sirenko2000, Fedorov1998, Ostapchuk2002} Notably, ferroelectricity near room temperature can be induced in STO films grown on DyScO$_3$ substrates which introduces a 1\% tensile strain.~\cite{Haeni2004, Skoromets2014} On the other hand, the 0.9\% compressive strain of a (La$_{0.3}$Sr$_{0.7}$)(Al$_{0.65}$Ta$_{0.35}$)O$_{3}$ (LSAT) substrate causes a sizeable hardening of the soft mode and a corresponding reduction of the dielectric constant.~\cite{Katayama2008} It has also been observed that the strain relaxation, which depends on the thickness of the STO films, and annealing can strongly affect the dielectric properties.~\cite{He2005, Zhai2009, Katayama2008, Kinjo2013} It is commonly assumed that the intrinsic strain condition can be maintained for ultrathin film with a thickness of about 10~nm. However, for such thin films, the direct observation of the soft mode in the far-infrared and terahertz range is a very difficult task due to limited sensitivity of conventional transmission and reflection techniques.~\cite{OHara2012, Tyunina2009, Cen2012}

Terahertz spectroscopic data have been previously reported on STO films on various substrates (Refs.~\cite{Fedorov1998, Ostapchuk2002, Kadlec2009,Katayama2008,Nuzhnyy2011, Katayama2008, Katayama2012, Kinjo2013, Kinjo2012}) with thicknesses as low as 17~nm,~\cite{Nuzhnyy2011, Fedorov1998} but typically on the order of hundreds of nanometers. The capability of far-infrared and terahertz ellipsometry to detect the soft mode in STO thin films has already been demonstrated,~\cite{Sirenko2000,Matsumoto2011} but so far no THz spectra were reported on ultrathin ($\sim 10$~nm) films. 
    
We show in the following that the combination of continuous wave Fourier-transform (FTIR) ellipsometry in the far infrared range~\cite{Bernhard2004, Hofmann2006, Kanehara2014} with time-domain terahertz (TD-THz) ellipsometry~\cite{Nagashima2001, Matsumoto2009, Matsumoto2011, Neshat2012, Morris2012} enables one to study the temperature evolution of the STO soft mode in thin (82~nm) and even ultrathin (8.5~nm) films and to obtain experimental quantities that can be directly interpreted in terms of the optical conductivity of the film.

The 82~nm and 8.5~nm thick SrTiO$_3$ films were grown by pulsed laser deposition (PLD) on (La$_{0.3}$Sr$_{0.7}$)(Al$_{0.65}$Ta$_{0.35}$)O$_{3}$ (LSAT) substrates with dimension of 10$\times$10$\times$0.5~mm$^3$ and 10$\times$10$\times$3~mm$^3$ (purchased from Crystec), respectively. They were deposited at 900~$^{\circ}$C in a partial pressure 0.12~mbar of O$_2$ with a laser fluence of 1.4~J/cm$^2$ and a repetition rate of 2 Hz. The growth was monitored with in-situ reflection high electron diffraction (RHEED). Bright RHEED patterns were observed that confirm the smoothness of the film surface. After deposition, the samples were cooled to 700~$^{\circ}$C at a rate of 10~$^{\circ}$C/min, while the oxygen partial pressure was gradually increased to 1 bar. Subsequently, the temperature was further decreased to room temperature at a rate of 5~$^{\circ}$C/min before the sample were removed from the PLD chamber. The thickness of the films was determined ex-situ with ellipsometry in the visible range using a Woollam VASE system. Both samples were measured before (as deposited) and after annealing by far-infrared and TD-THz ellipsometry. The ex-situ annealing was performed in ambient atmosphere and the cycle was equal for both samples: After an initial ramp, the temperature was maintained at 300~$^{\circ}$C for 2 hours, then it was increased during 4.5 hours to 1200~$^{\circ}$C where it was kept for 12 hours. The following cool-down from 1200~$^{\circ}$C to 200~$^{\circ}$C took 6 hours.

\begin{figure}
\vspace*{0cm}
\hspace*{0cm}
\includegraphics[width=6cm]{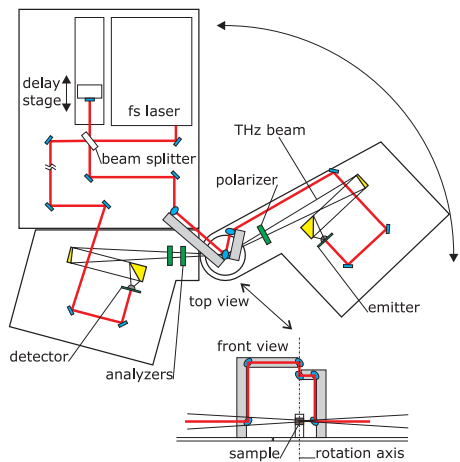}
\vspace{0cm}

\caption{\label{fig1SchemeSetup}
Schematic top view of the time-domain THz ellipsometer. The sample is located in the axis of rotation of the emitter arm. The front view shows the sample on the cold finger of the cryostat and the geometry of the laser-path bridge between the fixed part and the movable emitter arm.}
\end{figure}  

The far-infrared ($75-700$~cm$^{-1}$, $2.2-21$~THz) ellipsometry measurements were performed using a home-build ellipsometer attached to a Bruker IFS 113v spectrometer, similar to setup described in Ref.~\cite{Bernhard2004}, with a high-pressure Hg-lamp as light source and a 1.2~K Silicon bolometer detector. This ellipsometer operates in a rotating analyzer (RAE) configuration, with an optional reflection Si-prism compensator that has been used for the range of $75-165$~cm$^{-1}$ ($2.2-5$~THz).~\cite{Bernhard2004, Kanehara2014, Hoshina2014}

For the THz range ($3-70$~cm$^{-1}$, $0.1-2.1$~THz) we used a home-built time-domain THz ellipsometer, that is based on a Menlo Systems C-Fiber 780~nm pulsed laser, with 85~mW  output power and 100~fs pulse duration. Figure 1 shows a scheme of the setup. As THz source we used an area emitter with interdigitated electrodes (LaserQuantum Tera-SED)~\cite{Dreyhaupt2005} for which the polarization is oriented in the horizontal plane i.e. in the plane of incidence. The detector is a two-contact LT-GaAs photo-conductive antenna (Menlo systems TERA8-1) that is also oriented in the plane of incidence. The emitter arm can be rotated to obtain angles of incidence ranging from 45$^{\circ}$ to 90$^{\circ}$. To avoid the need of a realignment of the free space path of the pump laser, we incorporated a geometry where the fixed part of the path is coupled to the movable arm along the axis of rotation of the emitter arm, as sketched in the fig. 1. The polarizer and both analyzers are tandems of free standing wire grids (Specac GS57201). The ellipsometers (far-infrared and TD-THz) are equipped with a He-flow optical cryostat (from CryoVac). 
	
The far-infrared ellipsometer operates in the conventional continuous-wave mode with intensity detection. The analysis is performed with the standard formalism described in Refs.~\cite{bookTompkins, Bernhard2004, Kanehara2014}. Time domain spectroscopy, on the other hand, benefits from the ability to obtain the amplitude as well as the phase of wave arriving at the detector.~\cite{bookDexheimer, Grischkowsky1990} TD-THz ellipsometry was initially proposed to overcome the reference problem of reflection spectroscopy~\cite{Jeon1998, Pashkin2003} and several configurations and approaches have been discussed in the literature.~\cite{Nagashima2001, Khazan2001, Ino2004, Morikawa2006, Matsumoto2009, Neshat2012, Morris2012, George2012} We adopted the scheme used in FTIR RAE, performing measurements of the full time domain signals for a discrete set of analyzer angles $A_j$ (in our case $A_j=j\cdot36^{\circ}$, with $j=0$, 1, 2, 3, 4). A second fixed analyzer is required to maintain a constant polarization state of the pulse that is incident on the detector antenna.~\cite{Morikawa2006}

In the case of an isotropic sample with Jones matrix~\cite{bookTompkins} elements $r_{xx}$ and $r_{yy}$, the resulting ellipsometric ratio $\rho=r_{xx}/r_{yy}$ is usually expressed in terms of the ellipsometric angles $\Psi$ and $\Delta$. In the following we use instead a notation in terms of the complex pseudo-dielectric function:
\begin{equation}
\left\langle \varepsilon  \right\rangle ={{\sin }^{2}}\varphi \left[ 1+{{\tan }^{2}}\varphi {{\left( \frac{1+\rho }{1-\rho } \right)}^{2}} \right],
\end{equation}
where $\varphi$ is the angle of incidence. As will be outlined in the following, here the changes of $\langle\varepsilon\rangle$ with respect to bare substrate $\langle\varepsilon_{s}\rangle$ are proportional to the optical conductivity of the film. This can be seen by considering an ultrathin film for which $\lambda\gg{n}\cdot{d}$, with the wavelength of the THz radiation, $\lambda$, the film thickness $d$, and the refractive index, $n$ (with $\varepsilon=N^2=(n+ik)^2$). The optical response of the ambient/film/substrate system is described with the standard transfer matrix formalism.~\cite{bookTompkins} In the ultrathin limit, the propagation factors in the film, exp$(\pm{i}\omega\kappa{d}/c)$ , can be approximated as $1\pm{i}\omega\kappa{d}/c$, with $\kappa=\sqrt{\varepsilon-\sin^2\varphi}$  being the dimensionless normal component of the wave-vector in the film. By omitting terms that are quadratic in $d$, one obtains analytical expressions for $r_p$, $r_s$ and $\rho$ that can be inserted into eq. (1), to obtain
\begin{equation}
\left\langle \varepsilon  \right\rangle \doteq \left\langle {{\varepsilon }_{s}} \right\rangle -i\omega (\varepsilon -{{\varepsilon }_{s}})\cdot \frac{2d{{k}_{s}}}{c}\cdot \frac{\varepsilon -1}{\varepsilon }\cdot \frac{{{\varepsilon }_{s}}}{{{\varepsilon }_{s}}-1},
\end{equation}
where $\langle\varepsilon_{s}\rangle$ is the dielectric function of the substrate and $k_s=\sqrt{\varepsilon_s-\sin^2\varphi}$  the dimensionless normal component of the wave-vector in the substrate.~\cite{Aspnes1982}

\begin{figure}
\vspace*{0cm}
\hspace*{-0.4cm}
\includegraphics[width=9.3cm]{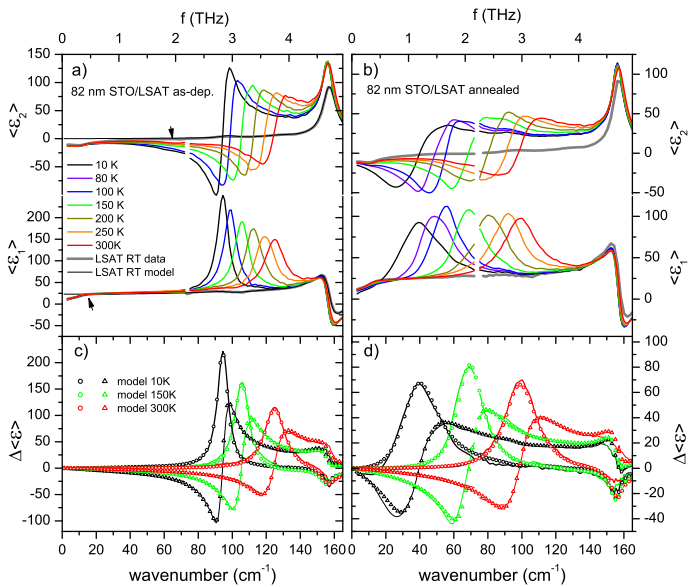}
\vspace{0cm}

\caption{\label{fig2STO_85nm}
Pseudo-dielectric functions of the 82~nm thick STO film on LSAT in the as deposited state (a) and the annealed state (b). The data of the bare LSAT substrate are shown by the gray line. The substrate and the film/substrate data deviate from the parameterized model (thin black line in a) below about 60~cm$^{-1}$ for $\langle\varepsilon_{2}\rangle$, and about 15~cm$^{-1}$ for $\langle\varepsilon_{1}\rangle$, as marked by arrows. Difference spectra of $\langle\varepsilon\rangle$ of the as-grown (c) and the annealed (d) film/substrate sample with respect to the substrate  at 10~K, 150~K and 300~K (lines), together with differential model spectra (circles for $\Delta\langle\varepsilon_{1}\rangle$, triangles for $\Delta\langle\varepsilon_{2}\rangle$). The color legend applies for all panels. 
}
\end{figure}

The second term contains the response of the thin film and it is proportional to the difference between the dielectric functions of the film and the substrate. The first pre-factor -$i\omega$  (except for a missing factor $\varepsilon_0$) converts the difference of $\varepsilon-\varepsilon_s$ to the one of the optical conductivities, $\sigma-\sigma_s$, where $\sigma_s$ is the optical conductivity of the substrate, i.e. $\Delta\sigma=\sigma-\sigma_s=-i\omega\varepsilon_0(\varepsilon-\varepsilon_s)$. The second factor is given by the film thickness, $d$, scaled by $k_s$. The factor $(\varepsilon-1)/\varepsilon$ represents the dielectric contrast between the film and the ambient with $\varepsilon_A=1$ (it becomes zero for $\varepsilon=1$ and yields a singularity at $\varepsilon=0$ that is known as the Berreman mode). Finally, the factor $\varepsilon_s/(\varepsilon_s-1)$ gives the inverse contrast between the ambient and the substrate, it leads to a singularity at $\varepsilon_s=1$ which gives rise to a surface guided wave.~\cite{bookSchubert, Dubroka2010, Berreman1963} Since the LSAT substrate is transparent in the THz range ($\varepsilon_s\approx23$), the factor $\varepsilon_s/(\varepsilon_s-1)$ is about 1.05 and $k_s\approx$4.7 for $\varphi=75^{\circ}$. The large value of the dielectric function of the STO film yields $(\varepsilon-1)/\varepsilon\approx1$ and gives rise to a sizeable contrast in $\Delta\sigma$. For example, a conductivity of the STO film of $\sigma_1$=100~$\ohmcm$ at a thickness of $d$=10~nm yields $\Delta\langle\varepsilon_{1}\rangle$=0.37. Note that eq.(2) is only weakly dependent on the angle of incidence $\varphi$ through $k_s$. The highest sensitivity (i.e. the minimum error of $\langle\varepsilon\rangle$) is obtained at the Brewster angle for $\varphi=\varphi_B$ which in case of LSAT with $\varphi_s$=23 is $\varphi_B=78.2^{\circ}$. The ellipsometric measurements presented here have been performed at $\varphi=75^{\circ}$.

This approach is especially useful in the spectral ranges where the response of the substrate is featureless. For the LSAT substrate this is the case well below a strong infrared active phonon mode at 158~cm$^{-1}$. Figures 2a and 2b clearly show the features of the soft mode of STO which gives rise to a characteristic peak in $\langle\varepsilon_1\rangle\sim\sigma_1$ and a corresponding resonance feature in $\langle\varepsilon_2\rangle\sim\sigma_2$ that is shifted to lower frequency at low $T$.
	
The spectra still contain some features that arise from the polarization dependent diffraction of the light due to the limited sample size. Such effects have been previously discussed for the case of bulk metallic samples in Ref.~\cite{Humlicek2004} These diffraction effects occur already in the response of the bare LSAT substrate (gray line in figures 2a, b) and can be clearly seen in comparison with the calculated THz response that has been obtained by modelling the phonon modes in the infrared regime with Lorentzian oscillators (thin black line). The measured data deviate from the model below about 60~cm$^{-1}$ where $\langle\varepsilon_2\rangle$ starts to bend down to negative values. In $\langle\varepsilon_1\rangle$ these diffraction effects give rise to a relatively sharp downturn below about 15~cm$^{-1}$, as marked by an arrow. Corresponding features are observed in the response of the film/substrate sample. 
	
These diffraction effects prevent us from analyzing and modelling the measured spectra of $\langle\varepsilon\rangle$. However, since the diffraction effects in the response of the bare substrate and the film/substrate sample are very similar, these are essentially absent in the difference spectra. This is shown in Figs. 2c and d in terms of the difference plots $\Delta\langle\varepsilon(T)\rangle= \langle\varepsilon(T)\rangle- \langle\varepsilon_s(T)\rangle$ for the 82~nm thick STO film in the as grown and the annealed state, respectively. As shown by the symbols, the dielectric response due to the soft mode of the STO film can now be well described with a Lorentzian oscillator model. The model calculates the pseudo-dielectric function of the film/substrate sample exactly, using the transfer matrix formalism, and subtracts the model dielectric function of the substrate. Note that the structure near 155~cm$^{-1}$ is a remnant of the strong phonon mode of the LSAT substrate, which, according to eq. (2), modifies the contrast between the film and the substrate.

The soft mode of STO is well accounted for by a single Lorentzian oscillator $\varepsilon (\omega )={{\varepsilon }_{\infty }}+4\pi {{F}^{2}}\omega _{0}^{2}/(\omega _{0}^{2}-{{\omega }^{2}}-i\omega \gamma )$. For the as grown 82~nm thick STO film the modelling yields a relatively sharp soft mode which at 300~K is centered at 126~cm$^{-1}$ with a width, $\gamma$, of about 20~cm$^{-1}$. With decreasing temperature it exhibits a fairly weak softening to 95~cm$^{-1}$ at 10~K. In the annealed state the soft mode frequency is considerably lower and the $T$-dependent softening is much stronger, i.e. the eigenfrequency decreases from about 100~cm$^{-1}$ at 300~K to about 40~cm$^{-1}$ at 10~K. The soft mode is also considerably broader than in the as grown state with $\gamma\sim$30~cm$^{-1}$ at 300~K and 35~cm$^{-1}$ at 10~K.

In the case of the as grown 82~nm film, below 100~K, two oscillators with similar eigenfrequencies $\omega_0$ but different $\gamma$ values improve the fit quality. This inhomogeneous broadening of the soft mode seems related to the antiferrodistortive phase transition from a cubic to a tetragonal structure which occurs in bulk SrTiO$_3$ at 105~K.~\cite{Lytle1964} It leads to the formation of domains with different orientations of the tetragonal axis and thus to strain fields in the vicinity of the domain boundaries which can affect the local soft mode behavior.
 
The strong reduction of the soft mode frequency due to a high temperature annealing treatment is in agreement with previous reports from THz transmission measurements of a 370~nm thick STO film on LSAT.~\cite{Katayama2008} Possible explanations are in terms of a relaxation of the strain state in large parts of such thick films or a removal of defects and vacancies that are also known to cause a hardening of the STO soft mode.

\begin{figure}
\vspace*{0cm}
\hspace*{0cm}
\includegraphics[width=5cm]{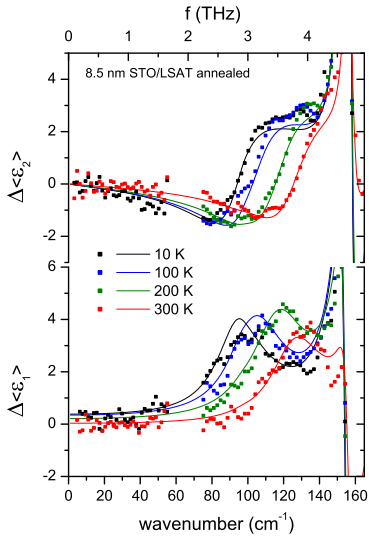}
\vspace{0cm}

\caption{\label{fig3STO_9nm}
Differential pseudo-dielectric function of the 8.5~nm thin STO film with respect to the bare LSAT substrate. 
}
\end{figure}

This raises the question whether a similar annealing-induced effect on the soft mode occurs in epitaxial STO films which are too thin to exhibit a large strain relaxation and thus remain clamped to the substrate. Thanks to the high sensitivity of the differential approach discussed above, we were able to observe the soft mode behavior in a 8.5~nm thick STO film on LSAT. To the best of our knowledge, this is the first direct observation of the soft mode for such an ultra-thin STO film. 

The as-deposited sample was studied with ellipsometry only at room temperature (data not shown). After the high temperature annealing treatment, the THz and infrared ellipsometry measurements have been performed as a function of temperature as shown in Figure 3 in terms of the differential data. Despite the much smaller signal, the characteristic features of the soft mode, such as the peak in $\Delta\langle\varepsilon_1\rangle\sim\sigma_1$ and the resonance feature in $\Delta\langle\varepsilon_2\rangle\sim\sigma_2$, are still clearly discernible. It is also evident that the dispersion of the soft mode is rather weak, i.e. it softens from about 128~cm$^{-1}$ at 300~K to 96~cm$^{-1}$ at 10~K. In comparison, in the as-deposited state the soft mode frequency at 300~K was at 132~cm$^{-1}$. For this ultrathin and fully strained STO thin film the high temperature annealing treatment thus has hardly influenced the position of the soft mode, the broadening however, increased from about 20 to 37~cm$^{-1}$.

\begin{figure}
\vspace*{0cm}
\hspace*{0cm}
\includegraphics[width=8.5cm]{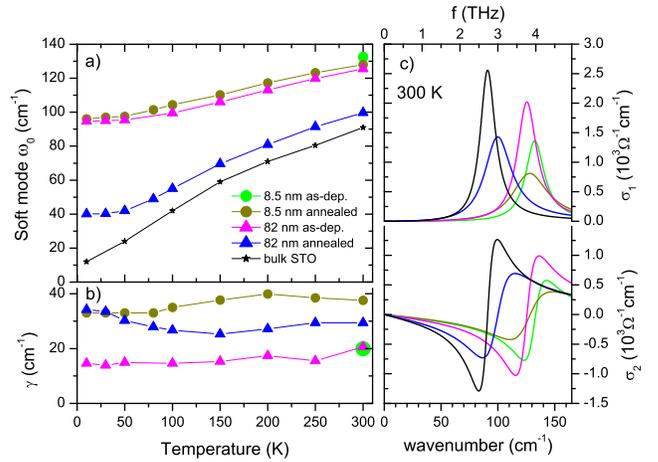}
\vspace{0cm}

\caption{\label{fig4Tdep}
Temperature dependence of the STO soft mode frequency, $\omega_0$ (a), and the broadening, $\gamma$ (b). Models fits of the optical conductivity at 300~K (c), with the same legend as a).   Bulk STO values from Ref.~\cite{Fedorov1998}.
}
\end{figure}

The overview of the temperature dependencies of the STO soft mode eigenfrequency and its broadening for the two STO films in the as grown and the annealed state is shown in Figures 4a and 4b, respectively. Also shown are literate data for bulk STO.~\cite{Fedorov1998} Figure 4c shows a comparison of the soft mode response at 300~K in terms of the modelled spectra. 
 
The good agreement between the absolute values and the temperature dependences of the soft mode frequencies in the annealed 8.5~nm and the as-deposited 82~nm STO films suggests that the strain relaxation in the latter is indeed rather weak. This conclusion should be further tested for example with x-ray diffraction experiments which can directly probe the relaxation of the lattice parameter. 
Independent from this open question, the presented work demonstrates the capability of the differential THz- and infrared ellipsometry technique to study the dielectric response of ultrathin STO layers. In the future, this technique can be used to study the strain effect on the soft mode behavior and subsequent ferroelectric transitions in fully strained films on various substrates.

In summary, we have shown that in the ultrathin film limit, the directly measured ellipsometric data – expressed in the form of pseudo-dielectric functions – can be readily interpreted as the sum of the dielectric function of the substrate (particularly in the range of its transparency) and the optical conductivity of the film. Both the film/substrate and the bare substrate data exhibit diffraction artefacts due to the limited sample size. These are removed in the differential pseudo-dielectric spectra, which can be subsequently analyzed with a differential model. We applied this approach on thin and ultrathin films of SrTiO$_3$ deposited on LSAT substrates and observed the temperature dependence of the soft mode dispersion. For the 82~nm thick STO film we observed a significant effect of the annealing treatment on the eigenfrequency of the soft phonon mode that was absent for the 8.5~nn film. This suggests that the strain due to the LSAT substrate can be readily maintained in such an ultrathin STO film.

\begin{acknowledgments}
This work has been supported by the Schweizer Nationalfonds (SNF) through the grant No. 200020-153660 and by the Sciex-NMS$^{CH}$ Grant No. CZ0908003. We acknowledge helpful discussions with Thomas Feurer, Andreas Bitzer, Florian Enderli, Gregory G\"{a}umann and Matthias R\"{o}ssle.
\end{acknowledgments}

\end{document}